\documentclass[fleqn,usenatbib]{mnras}

\usepackage[T1]{fontenc}

\usepackage{graphicx}   
\usepackage{amsmath}    
\usepackage{amssymb}    
\usepackage{booktabs}   
\usepackage{xspace}
\usepackage{ragged2e}

\newcommand{\cii}{[C\,{\sc ii}]}
\newcommand{\ci}{[C\,{\sc i}]}

\newcommand{\oiiil}{[O\,{\sc iii}] 88\,$\mu{\rm m}$}
\newcommand{\oiii}{[O\,{\sc iii}]}
\newcommand{\ciil}{[C\,{\sc ii}] 158\,$\mu{\rm m}$}

\newcommand{\zspec}{3.44\xspace}

\newcommand{\hncline}{HNC($J$\,=\,4--3)\xspace}
\newcommand{\cothreeline}{CO($J$\,=\,3--2)\xspace}
\newcommand{\cofourline}{CO($J$\,=\,4--3)\xspace}
\newcommand{\CIline}{[\ion{C}{i}]($^{3}P_{1}$--$^{3}P_{0}$)\xspace}

\newcommand{\Lsun}{L_{\odot}}
\newcommand{\Msun}{M_{\odot}}
\newcommand{\kms}{km\,s$^{-1}$\xspace}

\usepackage{newtxtext,newtxmath}
\definecolor{referee}{RGB}{0,0,0}

\definecolor{referee2}{RGB}{0,0,0}

\title[ALMA Band 2 of SPT0027-50 at $z=3.44$]{%
ALMA Band~2 line survey of a $z = 3.44$ clumpy strongly-lensed submillimetre galaxy}

\author[Bakx]{T. J. L. C. Bakx$^{1}$\thanks{E-mail: tom.bakx@chalmers.se}
\\
$^{1}$Department of Physics and Astronomy, Chalmers University of Technology, SE-412 96 Gothenburg, Sweden \\
}

\date{Accepted 2026 May 13. Received 2026 May 13; in original form 2026 April 02}

\pubyear{2026}

\begin{document}
\label{firstpage}
\pagerange{\pageref{firstpage}--\pageref{lastpage}}
\maketitle

\begin{abstract}
I present the first molecular line survey of the strongly lensed submillimetre galaxy {\color{referee} SPT-S\,J002706-5007.4} ($z = \zspec$) using the new Atacama Large Millimeter/submillimeter Array (ALMA) Band~2 receivers (67--116\,GHz), whose commissioning completes ALMA's full (sub-)millimetre frequency coverage. The broad spectral coverage from 76 to 111~GHz of the observations simultaneously accesses a large suite of molecular and atomic emission lines.
I report the novel detections of {\color{referee} the lines that were hitherto inaccessible at $z = 3.44$,} \cothreeline and \hncline, as well as detections of previously-observed \cofourline transitions, the neutral carbon line \CIline, HCN($J$\,=\,5--4), 
HCO$^{+}$($J$\,=\,5--4), and
HNC($J$\,=\,5--4), with fluxes in line with previous observations.  The CO spectral line energy distribution and \ci{}/CO line ratios indicate highly excited, dense molecular gas with a strong far-ultraviolet radiation field. The dense gas fraction is estimated at $17 \pm 9$\,per\,cent, consistent with other dusty star-forming galaxies selected from wide-area surveys. High-resolution Band~7 continuum imaging reveals a clumpy lensed morphology, with star-forming clumps contributing 30--50\,per\,cent of the total emission.
With multiple CO lines accessible across a wide redshift range, ALMA Band 2 is uniquely positioned as the premier tool for robust spectroscopic redshifts at Cosmic Noon and beyond ($z \sim 1$--$6$), a capability that will be further enhanced by the Wideband Sensitivity Upgrade's full-band coverage in fewer tunings.
\end{abstract}




\section{Introduction}
\label{sec:intro}
Fifteen years after the first light of the Atacama Large Millimeter/submillimeter Array (ALMA), the inclusion of its Band 2 (67--116\,GHz) heralds the completion of ALMA's spectral coverage \citep{Yagoubov2020}. As such, this completion can be rightly considered a moment of celebration for a large portion of the submillimetre and millimetre wavelength instrumentalists \citep{Huang2022,Claude2008,Kerr2014,Asayama2014,Belitsky2018,Ediss2004,Kerr2004,Mahieu2012,Sekimoto2008,Baryshev2015,Uzawa2013}, (submillimetre) observers and the rest of the astronomical community \citep[e.g.,][]{Mroczkowski2019}. To mark the occasion, several Band~2 Science Verification observations have been released to the community, and this manuscript focuses on the analysis of the high-redshift science case targeting one particular galaxy at $z = \zspec$ named SPT-S\,J002706-5007.4.

This galaxy is one of many now known submillimetre galaxies (SMGs; \citealt{Smail1997,Hughes1998,blain2002}) and/or dusty star-forming galaxies (DSFGs; \citealt{Casey2014}). 
Particularly those SMGs/DSFGs (hereafter DSFG for simplicity) discovered in wide-field submillimetre surveys represent some of the most vigorously star-forming systems in the Universe, with star-formation rates (SFRs) of order $10^{2}$--$10^{3}$\,$\Msun$\,yr$^{-1}$ \citep[e.g.][]{neri2020}. Their high abundance at Cosmic Noon ($z \sim 2$--$5$) implies that DSFGs dominate the contribution to the cosmic star-formation rate density (e.g., \citealt{Madau2014,Zavala2021}), and thus are key to understanding the build-up of most stars in the present-day Universe (e.g., \citealt{Tacconi2013}). The molecular gas reservoirs that fuel this prodigious activity can be characterised through millimetre and submillimetre spectroscopy \citep[e.g.,][]{Prajapati2026}, but the full complexity of the interstellar medium (ISM), i.e., its density, temperature, chemical enrichment, and dynamical state, requires observations of multiple tracers simultaneously \citep{Valentino2018,Rybak2020,Hagimoto2023}.

One of the largest surveys revealing dusty star-forming galaxies comes from the South Pole Telescope (SPT), as additional science to its main mission of characterizing the Cosmic Microwave Background radiation.
{\color{referee}The SPT has uncovered a large population of strongly lensed DSFGs across the southern sky \citep{Vieira2010,Everett2020,reuter20}}. Investigating these sources, and similar ones from surveys from the {\it Herschel Space Observatory} and {\it Planck} (e.g., \citealt{negrello2010,Negrello2014,negrello2017,wardlow2013,bussmann13}), have indicated that these sources are often amplified by gravitational lensing, which increases both the flux and apparent angular size of these objects, making them uniquely tractable targets for detailed molecular spectroscopy with ALMA. The target of the Band~2 scientific verification observations towards high redshifts is 
SPT0027-50. With its full name SPT-S\,J002706-5007.4 located at RA~=~00:27:06.54, Dec. = $-$50:07:19.8, it has a redshift close to the average of the SPT-galaxy sample with $z_{\rm spec} = \zspec$ (\citealt{reuter20}). The source is among the brightest of these lensed sources, with a lensing magnification factor of $\mu \simeq 5.5$ \citep{Spilker2016,reuter20} and an intrinsic far-infrared (FIR) luminosity of
$L_{\rm FIR} \simeq 8.5 \times{}10^{13}$\,$\Lsun$, corresponding to a SFR of 3200\,$\Msun$\,yr$^{-1}$.

\begin{figure*}
    \centering
    \includegraphics[width=\linewidth]{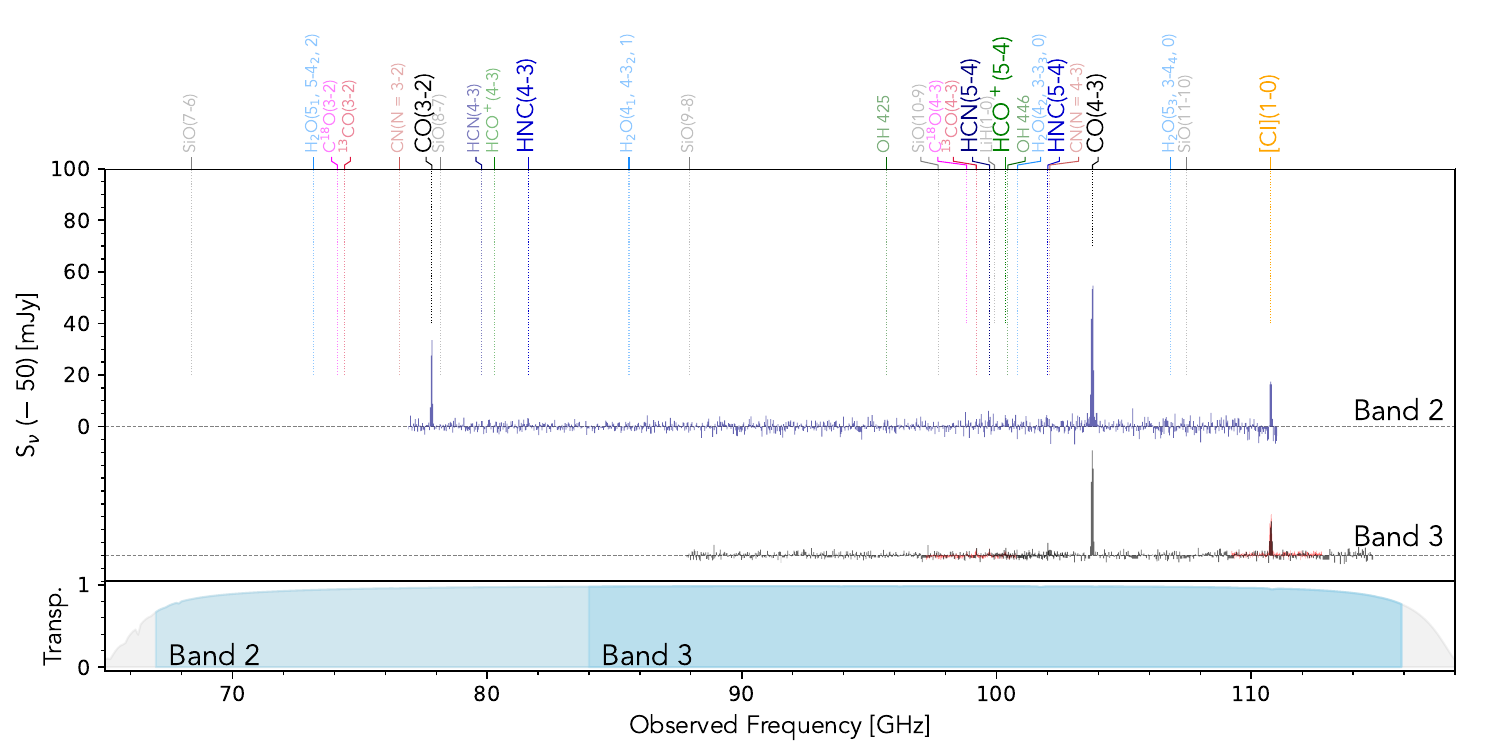}
    \caption{{\color{referee2} The Band 2 and 3 spectra, with the Band~2 data shown as a \textit{thin blue step graph}, the Band~3 data from 2022.1.00172.S and 2022.1.00526.S shown as an offset \textit{red step graph}, and the Band~3 data from 2015.1.00504.S shown as an offset \textit{wide grey step graph}.} Extending beyond the Band~3 regime, the Band~2 data provide a deep view of spectral lines previously not available to ALMA, including the CO(3--2), HCN(4--3), HCO(4--3) and HNC(4--3) lines. The top part of the figure indicates bright emission lines (from~\citealt{Spilker2014,Hagimoto2023}) in the range of ALMA Band~2, with smaller fonts indicating undetected lines, while the larger fonts show the detected lines shown in detail in Figure~\ref{fig:linespectra}. The bottom part shows, for good observing conditions (precipitable water vapour of 0.5~mm), the atmospheric transparency and observing windows of ALMA Band~2 and Band~3. }
    \label{fig:spectrum}
\end{figure*}

Here, I report first results from our ALMA Band~2 observations of SPT0027-50, and contextualize them within larger studies of dusty star-forming galaxies. In Section~\ref{sec:obs}, I discuss the observations and data reduction and fluxes found in Band~2, {\color{referee2} as well as supplemental Band~3 ($\sim 100$~GHz) and Band~7 ($\sim 343$~GHz)} archival data. In Section~\ref{sec:disc}, I identify the source nature and conclude with the prospects of future surveys with Band~2 including redshift searches with Band~2 and others, especially in the Wideband Sensitivity Upgrade era in Section~\ref{sec:zsearches}.
Throughout this paper, I assume a flat $\Lambda$-CDM cosmology with the best-fit parameters derived from the \textit{Planck} results \citep[][paper VI]{Planck2020}, which are $\Omega_\mathrm{m} = 0.315$, $\Omega_\mathrm{\Lambda} = 0.685$ and $h = 0.674$.

\section{Observations, Data Reduction, and Fluxes}
\label{sec:obs}
{\color{referee} With Band~2 available across 25 telescopes, ALMA targeted SPT0027-50 in Spectral Scan mode in Band~2 with a total of twenty 1.875GHz spectral windows covering from approximately 76.87 to 111.03 GHz (2011.0.00025.SV; P.I. M. D\'iaz-Trigo). SPT0027-50 was selected as a verification target because it is one of the brightest $z > 3$ sources in the SPT sample. It has several bright spectral lines, including CO(3--2), at frequencies that before Band~2 were not available to ALMA, and these Band~2 observations also cover a large inventory of lines including HCN, HNC and HCO$^{+}$ to do a scientific exploration of dense gas inside this bright dusty galaxy. These verification observations thus offer a unique possibility to showcase the lower-frequency potential now opened up by Band~2 \citep{Mroczkowski2019}, in particular due to Band~2's larger Intermediate Frequency (IF) range, which facilitates a contiguous stacking of spectral windows useful in line surveys including redshift searches.} The observations were carried out on {\color{referee} 2025 November 27} within configuration C43-6, with telescope baselines ranging from 15.3 m to 2.4 km. 

Calibration and imaging were performed using the ALMA Pipeline, with the calibrated files made available by the ALMA scientific support team\footnote{\url{https://almascience.eso.org/alma-data/science-verification/spt0027_band2}}. Auto-masking was used for generating the images unless indicated otherwise. 
All Band~2 data were processed using the \textsc{Common Astronomy Software Applications} (\textsc{CASA}) package \citep{McMullin2007,casateam2022}, using CASA version 6.5.6. After this, the data were imaged using
\textsc{tclean} with a robust parameter of 2 (i.e,. natural weighting). The resulting per-bin sensitivity of the data cubes is between 0.73 and 1.33 mJy per beam in 35~km~s$^{-1}$ bins with beam sizes ranging from 0.62 by 0.47~arcseconds to 0.87 by 0.76~arcseconds. Note that in the $\sim 0.2$~GHz overlapping frequency regions between two spectral windows, the Band~2 observations are able to achieve a sensitivity of $\sim 0.63$~mJy per beam. 
The continuum data has a per-beam sensitivity of 5.6~$\mu$Jy per beam and a beam size of 0.72 by 0.60~arcseconds. 

For comparison, I also collect additional publicly-available Band~3 ALMA spectroscopy (2015.1.00504.S; P.I. M. Strandet, 2022.1.00172.S; P.I. C. Yang, and 2022.1.00526.S; P.I. A. Weiss), as well as Band~7 data from a higher-resolution imaging (2023.1.01585; P.I. J. McKean). 
For the Band~3 2015.1.00504.S project, imaging and spectroscopy was taken directly from the ALMA archive, consisting of eight data cubes together contiguously covering 87.5 to 114.75~GHz with sensitivities between 1.7 and 1.0~mJy per beam and a beam size of 4.85 by 4.52~arcsec. 
For all other publicly-available data, I used the provided \textsc{scriptForPI.py} to create calibrated data files. After this, the data were imaged using
\textsc{tclean} with a robust parameter of 2 (i.e,. natural weighting). 
Covering 85.6 to 88.5 and 96.7 to 100.4~GHz, 
the 10~km~s$^{-1}$ sampled data cubes from the Band~3 project 2022.1.00172.S have a per-beam sensitivity of 0.4~mJy per beam with a beam size of 1.25 by 1.02~arcseconds. 
Covering 97.2 to 100.8 and 109.2 to 112.8~GHz, the 35~km~s$^{-1}$ sampled data cubes from the Band~3 project 2022.1.00526.S have a per-beam sensitivity of 0.25~mJy per beam with a beam size of 1.5 by 1.4~arcseconds. Covering 335.9 to 339.7 and 347.9 to 351.8~GHz, the Band~7 project imaging 2023.1.01585.S has a per-beam sensitivity of 38~$\mu$Jy per beam with a beam size of 50 by 44~milliarcseconds.

Figure~\ref{fig:spectrum} shows the Band~2 and Band~3 spectra for SPT0027-50. 
The spectra are extracted from each data cube using a single wide aperture that extends down to two sigma in the Band~2 dust continuum, and is widened by two beams. Together with the spectra, the indications of expected emission lines are shown as dashed lines, with Band~2 detected lines (see below) indicated in larger font. 

\begin{figure}
    \centering
    \includegraphics[width=\linewidth]{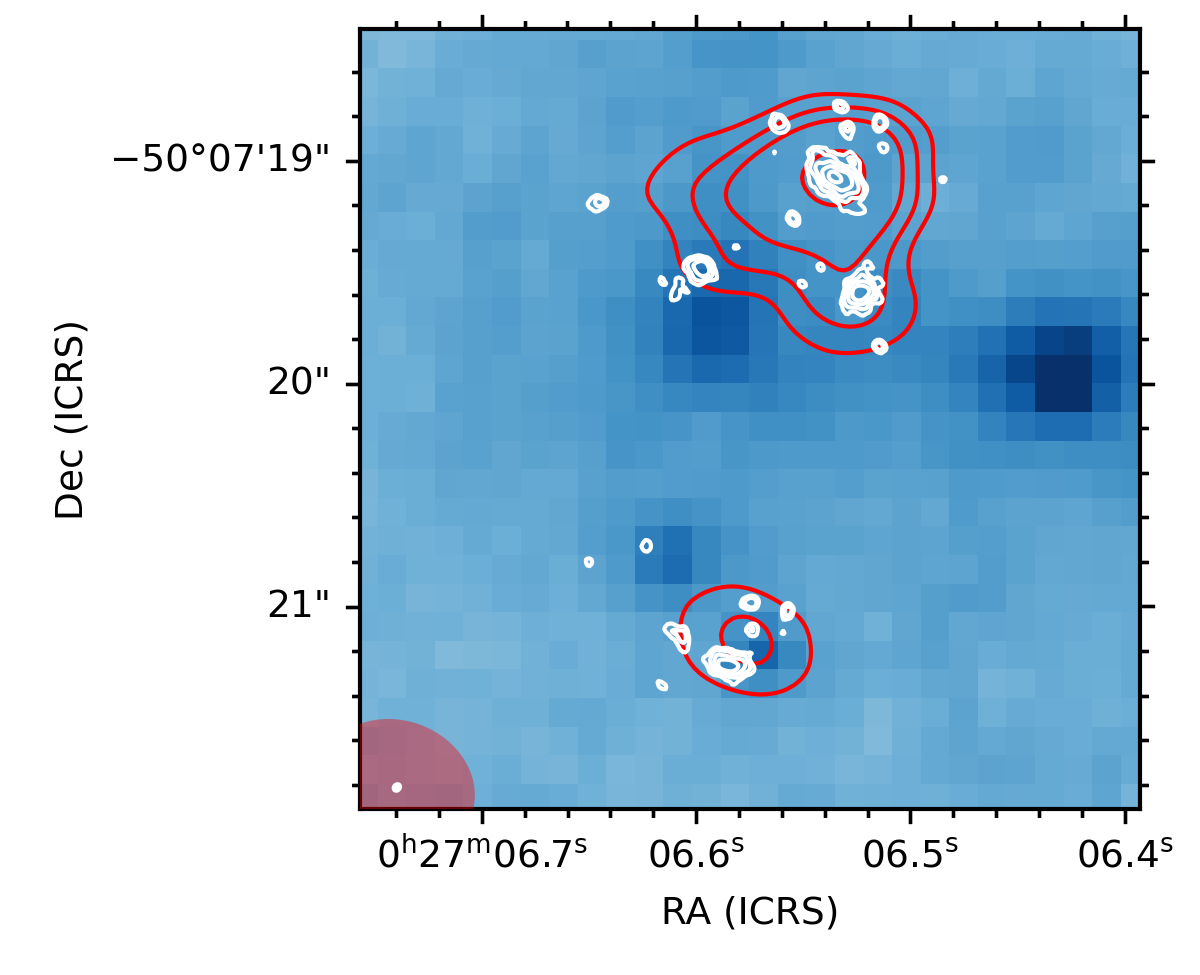}
    \caption{HST imaging on SPT0027-50 (background) with both the ALMA Band~2 continuum imaging (red thick contours) and higher-resolution Band~7 continuum imaging (thin white contours). The beams of each of the ALMA observations are shown in the bottom left. The lensing morphology of SPT0027-50 is suggested in the Band~2 imaging (as well as previous lower-resolution Band~7 analysis; \citealt{Spilker2016}), with the clumpy lensed nature of SPT0027-50 appearing more apparent in the higher-resolution imaging.}
    \label{fig:continuum}
\end{figure}

The Band~2 dust continuum map of SPT0027-50 is produced by averaging all line-free channels, and reveals a geometry consistent with the known lensing geometry of the system \citep{Spilker2016}. 
The integrated continuum flux density is
$S_{\rm Band\ 2} = 147 \pm 23$\,$\mu$Jy,
corresponding to an observed average frequency of 93.935\,GHz.
Figure~\ref{fig:continuum} shows the continuum map of SPT0027-50 in both Band~2 as well as the higher-resolution Band~7. {\color{referee} The background image is from publicly-available \textit{Hubble Space Telescope} imaging (PID: 12659, PI: J. Vieira) previously reported in \cite{Spilker2016}. These observations target the observed-frame $\sim 1.6$~$\mu$m emission using the F160W filter of the WideField Camera 3 on the \textit{HST}.  }

\begin{figure*}
\includegraphics[width=0.495\linewidth]{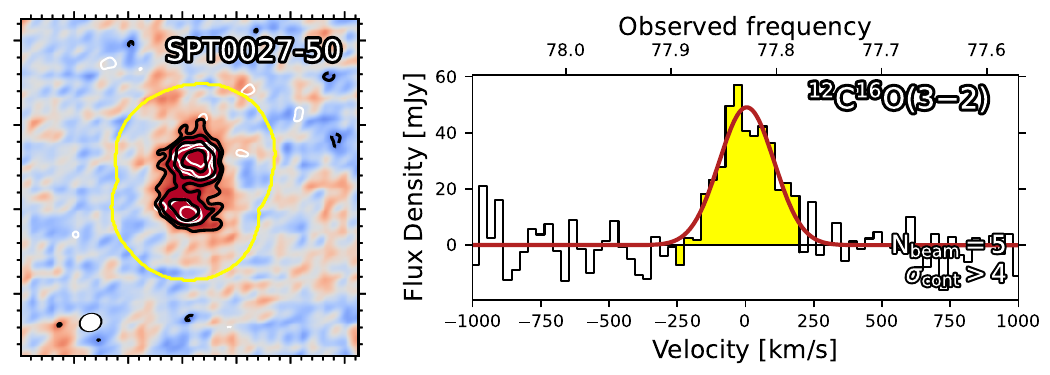}
\includegraphics[width=0.495\linewidth]{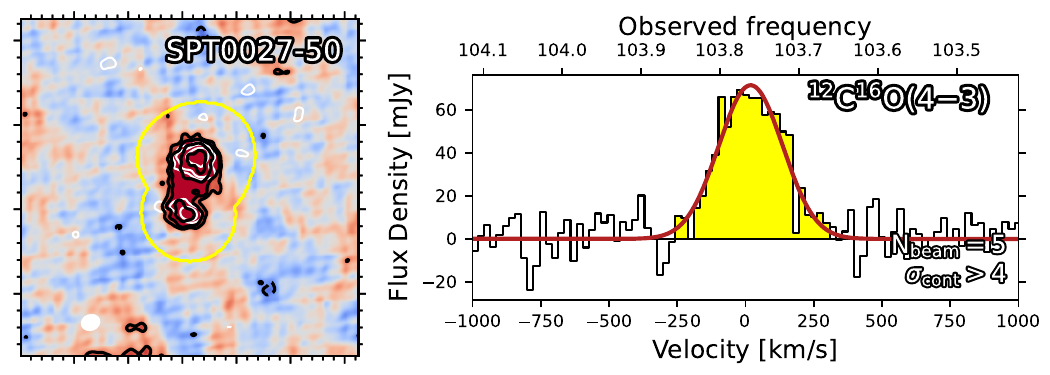}
\includegraphics[width=0.495\linewidth]{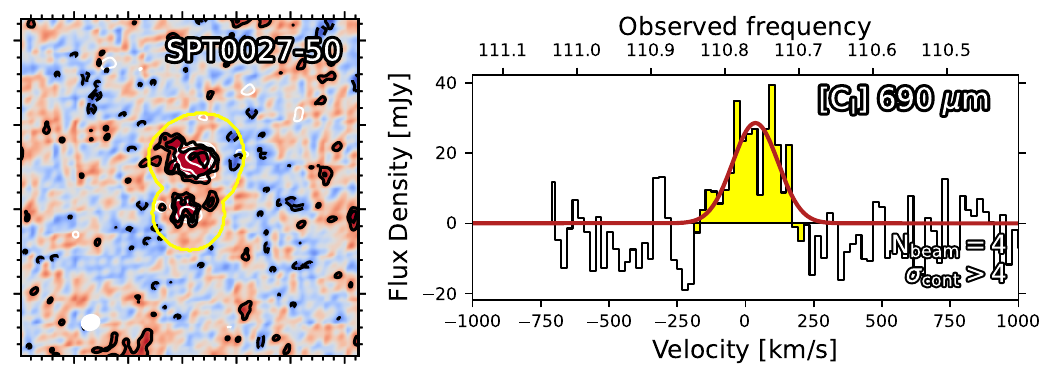}
\includegraphics[width=0.495\linewidth]{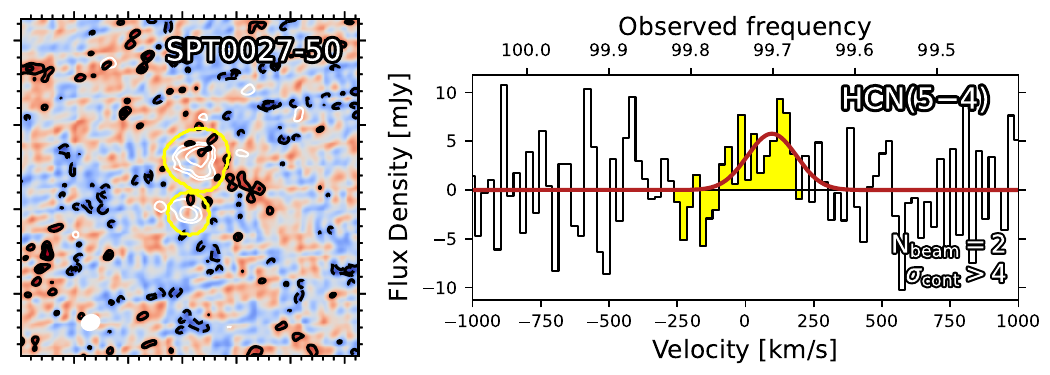}
\includegraphics[width=0.495\linewidth]{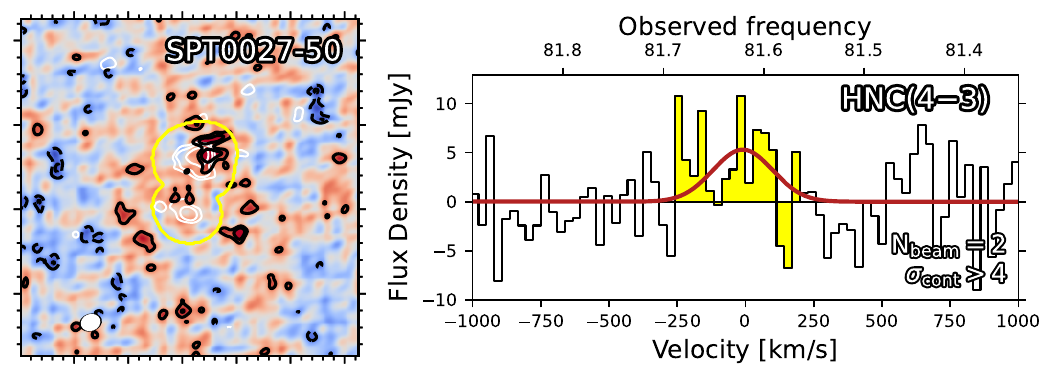}
\includegraphics[width=0.495\linewidth]{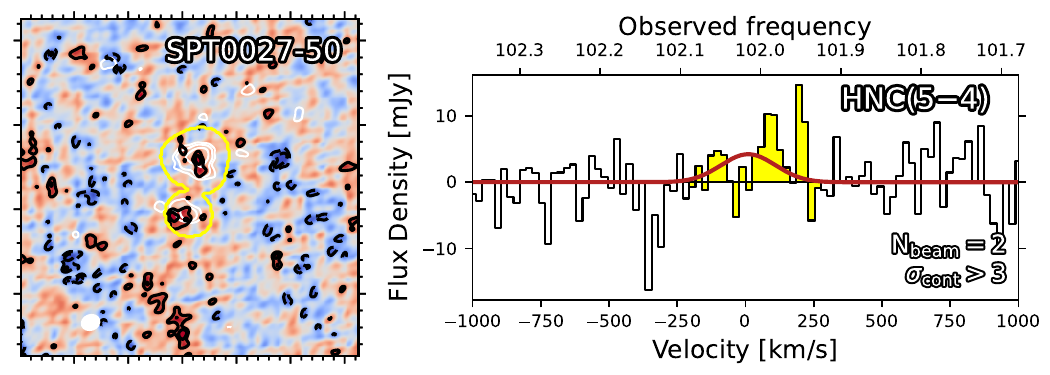}
\includegraphics[width=0.495\linewidth]{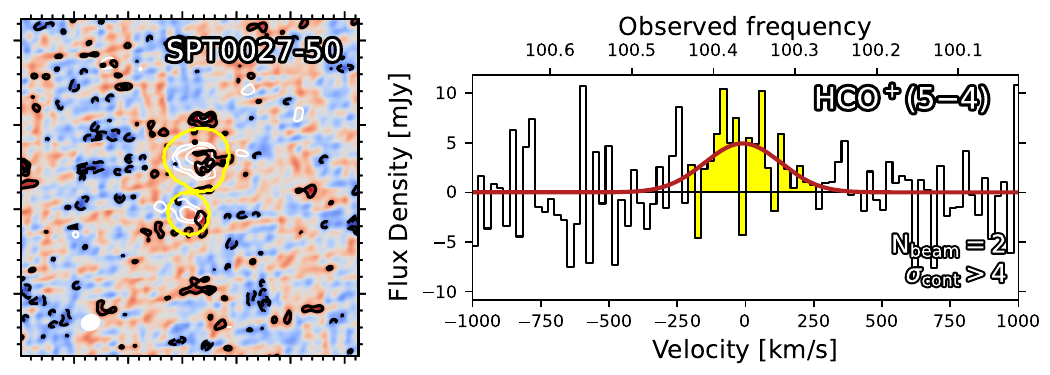}
    \caption{The spectral lines extracted from the Band~2 data. Per line, the left-hand side panel indicates the moment-0 map of the line (background and black thick contours), while the thin white contours indicate the continuum, and the yellow contours indicate the aperture used for the flux extraction. The right-hand side spectrum is shown as a function of velocity relative to the rest-frame, with the top $x$-axis indicating the observed frequency. The filled region indicates the velocity range used for generating the moment-0 map, and the bottom-right values for $N_{\rm beam}$ and $\sigma_{\rm cont}$ indicate the criteria for identifying the extraction contours in the data cube (i.e., the yellow contours in the left-hand side image).}
    \label{fig:linespectra}
\end{figure*}

Line fluxes are extracted using the procedure detailed in \cite{Bakx2024ANGELS} and \cite{Bakx2026} and are provided in Table~\ref{tab:lines}. I show the continuum-subtracted lines in Figure~\ref{fig:linespectra}. In short, an aperture is created that best matches the expected emission based on ancillary data. I use the ALMA Band~2 continuum image as a starting aperture for the line extraction, which is then smoothed to the resolution of the data cube. Using the average beam of the data cube, the aperture is widened by up to five beams if the signal-to-noise ratio permitted\footnote{Indicated by $N_{\rm beam}$ in Figure~\ref{fig:linespectra}; the extent of subsequent smoothing is decided on a per-source basis, ensuring no significant flux is missed but also minimizing the additional noise that comes from an aperture that is too large.} to ensure extended flux is also included in the flux measurement. The total measurement of the velocity-integrated line flux includes the errors in velocity and peak flux, and therefore its uncertainties do not reflect the significance of the line detection, which is always in excess of $4 \sigma$ in the moment-0 map.

Using the catalogue of bright lines expected for DSFGs from \citealt{Spilker2014,Hagimoto2023} and \citealt{Reuter2023}, I investigate all lines within the observed wavelengths. This proved successful for both CO lines (CO(3--2), CO(4--3)), the atomic Carbon line, \ci{} 609~$\mu$m, and four additional dense-gas tracers, namely 
HCN(5--4), 
HNC(4--3), 
HNC(5--4), 
and 
HCO$^+$(5--4). The HCN(4--3) and HCO$^+$(4--3) dense gas tracers, as well as the CO isotopologues $^{13}$CO(4-3), and C$^{18}$O(4-3), proved undetected in the Band~2 data. For completeness, they are added to Table~\ref{tab:lines} in order to compare them to ancillary data available on the ALMA archive. All line fluxes are within one combined standard deviation away from the additional detected measurements of lines from archival projects.

\begin{table}
    \centering
    \caption{Bright molecular and atomic emission lines within the ALMA Band~2
    window at $z = 3.44$.}
    \label{tab:lines}
    \begin{tabular}{lrrr}
        \hline
        Line & $\nu_{\rm rest}$ & $\nu_{\rm obs}$ & $S\Delta v$ \\ 
             & [GHz] & [GHz] & [Jy\,\kms] \\
        \hline
CO(3--2)			& 345.7960				& 77.8293	& 12.35 $\pm$ 1.45$^{\dagger}$ \\ 
CO(4--3)			& 461.0408				& 103.7679	& 21.05 $\pm$ 1.57$^{\dagger}$ \\ 
&&& 12.84 $\pm$ 1.11$^{*}$  \\
$^{13}$CO(4--3)			& 440.7652	& 99.2044				& $<$0.84$^{\dagger}$ \\ 
&&&$<$0.93$^{*}$ \\ 
&&&0.64 $\pm$ 0.10$^{\#}$\\
&&&0.43 $\pm$ 0.17$^{\P}$\\
C$^{18}$O(4--3)			& 439.0888	& 98.8270				& $<$2.01$^{\dagger}$ \\ 
&&&$<$1.53$^{*}$ \\ 
&&&0.46 $\pm$ 0.09$^{\#}$\\
&&&0.26 $\pm$ 0.14$^{\P}$\\
\ci{} 609~$\mu$m			& 492.1606			& 110.7721		& 5.96 $\pm$ 1.26$^{\dagger}$ \\ 
&&&4.07 $\pm$ 1.21$^{*}$ \\
&&&4.19 $\pm$ 0.25$^{\P}$\\
HCN(4--3)			& 354.5055			& 79.7896		& $<$2.01$^{\dagger}$ \\ 
HCN(5--4)			& 443.1161			& 99.7335		& 1.30 $\pm$ 0.69$^{\dagger}$ \\ 
& & & 2.18 $\pm$ 0.89$^{*}$ \\
&&&0.82 $\pm$ 0.12$^{\#}$\\
HNC(4--3)			& 362.6303			& 81.6183		& 1.45 $\pm$ 0.82$^{\dagger}$ \\ 
HNC(5--4)			& 453.2699			& 102.0188		& 1.32 $\pm$ 0.63$^{\dagger}$ \\ 
&&&1.85 $\pm$ 1.15$^{*}$\\
HCO$^+$(4--3)			& 356.7342	& 80.2912				& $<$1.23$^{\dagger}$ \\ 
HCO$^+$(5--4)			& 445.9029	& 100.3607				& 1.73 $\pm$ 0.89$^{\dagger}$ \\ 
&&&0.73 $\pm$ 0.47$^{*}$\\
&&&0.68 $\pm$ 0.12$^{\#}$\\
&&&0.38 $\pm$ 0.13$^{\P}$ \\
        \hline
    \end{tabular}
 { \raggedright \\    \textbf{Notes:} 
Col. 1: Line name
Col. 2: Rest-frame frequency.
Col. 3: Observed-frame frequency.
Col. 4: Velocity-integrated flux density for Band~2 and Band~3 observations, with the below markers indicating the respective origins.
$^{\dagger}$ Line fluxes from the Band~2 Science Verification project 2011.0.00025.SV.
$^{*}$ Line fluxes from the Band~3 project 2015.1.00504.S.
$^{\#}$ Line fluxes from the Band~3 project 2022.1.00172.S.
$^{\P}$ Line fluxes from the Band~3 project 2022.1.00526.S.
    }
\end{table}

\section{Discussion}
\label{sec:disc}

\subsection{ISM conditions from the Band~2 window}
\label{sec:ism}
Although many of the lines were already available from ancillary data, the Band~2 data provide an excellent inventory of gas tracers to characterise the ISM.
The set of molecular lines are simultaneously able to probe both the molecular and dense gas inside a galaxy at Cosmic Noon. 

The flux ratio between \cothreeline and \cofourline find a ratio of $1.7 \pm 0.3$, where I note that the Band~2 data allows us to compare these values directly through a singly-defined aperture. This ratio is close to the maximum expected line ratio based on a thermalized line profile, which is expected to rise as $J_{\rm up}^2$, with 1.77 ($= 16/9$) as a theoretical maximum ratio (see \citealt{Hagimoto2023} and \citealt{Serjeant24} for an extensive discussion of line ratios exceeding CO line thermalization). As such, the CO Spectral Line Energy Density (SLED) of SPT0027-50 indicates high ISM gas densities and kinetic temperatures \citep{Dannerbauer2009,Daddi2015A&A...577A..46D,Yang2023}.

The [\ion{C}{i}]($^{3}P_{1}$--$^{3}P_{0}$) fine-structure transition at 492.161\,GHz falls at an observed frequency of 110.772\,GHz, near the upper edge of the Band~2 window. Atomic carbon is a useful tracer of atomic gas mass complementary to the molecular gas tracing of CO, with the [\ion{C}{i}]/CO ratio providing a measure of the photodissociation region and the $\rm [C/H]$ abundance \citep{Kaufman2006}. Using the equations from \cite{Solomon2005}, the line luminosity ratio of \CIline/\cofourline is 0.30 $\pm$ 0.06, which is in line with gas densities ($n_{\rm H}$) of $10^{4.5}$ \citep{Kaufman2006,Hagimoto2023}. The line luminosity of $L_{\rm [C_I]}$ is (6.5 $\pm$ 1.4) $\times 10^8$~L$_{\odot}$, for a \ci{} line-to-infrared luminosity of $(1.38 \pm 0.27) \times 10^{-6}$, suggesting a relatively strong far-ultraviolet radiation intensity field of 10$^{3.8}$~Habing. 

The simultaneous coverage of the HCN($J$\,=\,4--3), HCO$^{+}$($J$\,=\,4--3), and HNC($J$\,=\,4--3) transitions trace molecular gas at higher densities $n_{\rm H_2} \gtrsim 10^{5}$--$10^{6}$\,cm$^{-3}$. Albeit not detected in HCN($J$\,=\,4--3) and HCO$^{+}$($J$\,=\,4--3) in this case, each line species is detected across the Band~2 spectral scan in their (5--4) transition, providing among the highest-redshift detections of these tracers reported to date \citep{Rybak2026}.

The line luminosity ratio of $L'_{\rm HCN(5-4)} / L'_{\rm CO(4-3)} = 0.067 \pm 0.036$ is relatively high when compared to stacks \citep{Spilker2014,Hagimoto2023,Reuter2023} but appears in agreement with results from samples of individual galaxies \citep{Rybak2026}. {\color{referee} The observed lines provide an opportunity to estimate the dense gas fraction of SPT0027-50 using the scaling relations $M_{\rm dense\ gas} = \alpha_{\rm HCN} r_{j,1} L'_{\rm HCN(j, j-1)}$ and $M_{\rm molecular\ gas} = \alpha_{\rm CO} r_{j,1} L'_{\rm CO(j, j-1)}$}. In these equations, the $\alpha$ value indicates the gas conversion coefficient, ad-hoc taken to be 10 $M_{\odot}$ (K km s$^{-1}$ pc$^2$)$^{-1}$ for HCN(1--0) and 1~$M_{\odot}$ (K km s$^{-1}$ pc$^2$)$^{-1}$ for CO(1--0) in \citet{Rybak2026}, and the respective $r_{j,1}$ line luminosity ratios between the observed transition and the ground state. Given the high $r_{4,3}$ ratio observed in the Band~2 observations, I assume $r_{4,1} = 1$ in line with highly excited gas \citep{Riechers2011QSO}, while for HCN I assume $r_{5,1} = 0.25$ extrapolated from the observations in \citet{Israel2023} and \citet{Rybak2026}. The subsequent ratio of dense to total molecular gas is thus $17 \pm 9$~per cent, although this value can two-fold increase for lower values of CO line luminosity ratios typically assumed for DSFGs \citep{Harrington2021}. This value appears in line with values of other DSFGs selected from large-area surveys \citep{Rybak2026}.

The HCN, HNC and HCO$^{+}$ line ratios are in agreement with 1, given the large uncertainties on individual values and the compounding effect of errors within ratios. As such, I find no evidence for significant effects of an Active Galactic Nucleus within the SPT0027-50 system \citep[e.g.,][]{Kohno2003,Krips2008,Imanishi2016}.
Finally, none of the CO isotopologues are directly detected in the Band~2 data, while archival data indicates isotoplogue ratios of between $^{12}$C/$^{13}$C $\approx$ 30 to $^{16}$O/$^{18}$O $\approx$ 50. These values are in agreement with previous studies of stacks \citep{Hagimoto2023}, and subsequent studies of this archival data can provide independent probes of the nucleosynthetic enrichment history of the ISM \citep{henkel2010,Zhang2018}, but exceed the scope of this manuscript given the Band~2 non-detections.


The exceptional frequency coverage of the ALMA Band~2 receivers now provides the opportunity for rich molecular line surveys of a high-redshift galaxies, in a frequency coverage previously only possible with NOEMA \citep[e.g.,][]{Yang2023} and single-dish observations \citep[e.g.,][]{zavala2018nat}.
The breadth of this dataset motivates future deeper observations in Band~2 with increased integration time to detect weaker lines, including: (i) higher-$J$ transitions of rare isotopologues to constrain isotopic ratios (see e.g., \citealt{Rybak2026}); (ii) complex organic molecules as tracers of warm chemistry; (iii) absorption of lines against the CMB \citep{Riechers2022}.
As such, this source and its Band~2 observations will serve as a benchmark for ALMA observations in the cycles to come.

\subsection{A clumpy galaxy with Band~2 observations}
Figure~\ref{fig:continuum} shows the spatial distribution of Band~7 emission from SPT0027-50. The cuspy lensed profile shows four images of the source, with significant emission coming from the clumps. The molecular and dense gas emission lines from Band~2 indicate an interesting source with a high gas density and radiation field intensity, and the clumpy nature warrants further characterization of this system within the larger interpretation of dusty galaxies \citep[e.g.,][]{Kamieneski2024,Bakx2024ANGELS}. By measuring the flux of the clumpy extended emission between the four images, and comparing this to the total emission, I find the clumps to contribute between 30 to 50~\% of the total emission of the source. This clumpy morphology of SPT0027-50 places it within a broader class of intensely star-forming galaxies whose internal structure has been the subject of considerable debate. 

Giant star-forming clumps, with masses of order $10^{8 - 9}$~$\Msun$, are thought to play a central role in bulge assembly, the regulation of star formation, and the growth of central black holes \citep{ForsterSchreiber2006,Elmegreen2008,Elmegreen2009}. Their formation is most naturally explained by gravitational fragmentation of gas-rich, turbulent discs with Toomre $Q < 1$ \citep[e.g.][]{Bournaud2007, Krumholz2012ApJ...745...69K}, though galaxy mergers have also been proposed as a contributing channel \citep{DiMateo2008, Calabro2019}. 
The submillimetre view of clumpiness has been particularly contested. Whilst resolved ALMA observations of some SMGs reveal clumpy dust-continuum and CO emission \citep[e.g.][]{Tadaki2018,Hodge2019}, others find remarkably smooth, centrally concentrated distributions \citep{Gullberg2019,Rujopakarn2019,ivison2020}, with the cold dust appearing smoother than the molecular gas owing to its more efficient cooling. In the handful of systems where clumpy CO emission is detected, the constituent giant molecular clouds lie systematically offset from local Larson scaling relations, exhibiting higher gas mass surface densities and larger internal velocity dispersions, yet remaining gravitationally bound, properties more akin to GMCs in local starbursts and mergers than to quiescent disc systems \citep{dessauges-zavadsky2019}. 

The clear clumpy morphology of SPT0027-50 therefore makes it a valuable counterexample to the smooth-dust scenario, and the gravitational lensing magnification, resolving structures on scales of order 150\,pc, offers a rare opportunity to study individual clumps in a $z \sim 3.5$ starburst at a level of detail otherwise inaccessible. Future high-resolution observations targeting CO and [C\,{\sc ii}] emission will be essential to determine whether these clumps are rotationally supported fragments of a turbulent disc or the product of a recent merger event.

\begin{figure*}
    \includegraphics[width = 0.31 \linewidth]{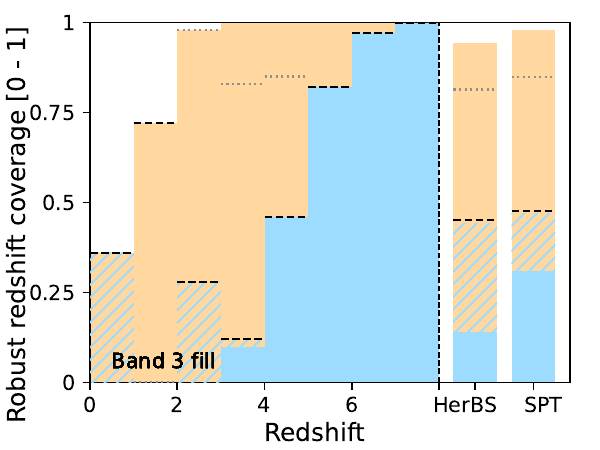}
    \includegraphics[width = 0.31 \linewidth]{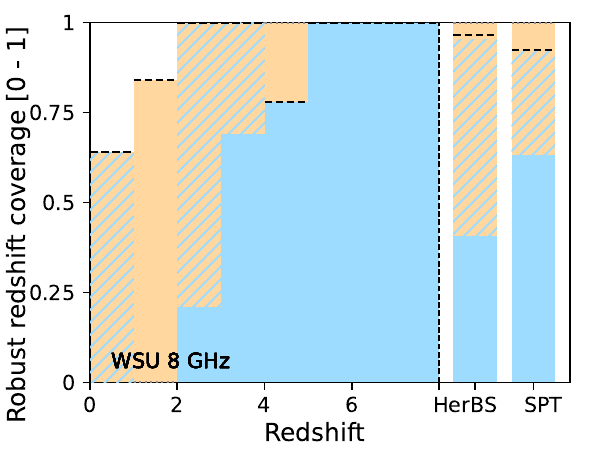}
    \includegraphics[width = 0.31 \linewidth]{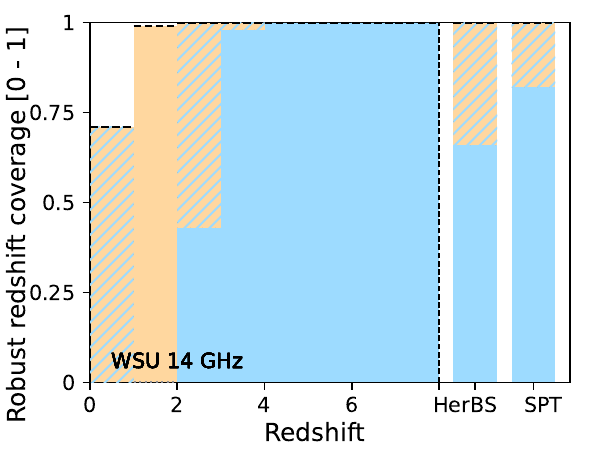}
\caption{Redshift identification probability as a function of redshift ($0 \leq z \leq 8$) for three ALMA frequency configurations, evaluated against smoothed redshift distributions from the HerBS \citep{bakx18, Bakx2020} and SPT \citep{reuter20} samples. Orange bars indicate the fraction of sources yielding a single detected spectral line, while blue bars indicate two or more detections. Hatched blue regions denote configurations where even a single line provides an unambiguous redshift identification; hatched orange regions indicate cases where redshift degeneracies persist given a 5$\sigma$ uncertainty in $z_\mathrm{phot}$. Grey dotted lines show the robust redshift fraction when \ci{} emission is included, and black dashed lines show the improvement expected from refined photometric redshift uncertainties \citep{Bendo2023}.
The three panels correspond to \textit{left}: the filled Band~3 tuning \citep{Weiss2013}, which serves as the baseline, \textit{middle and right}: the Band~2-based configurations that use the WSU with 8 (baseline) and 14~GHz (goal) bandwidths. For the 8~GHz set-up, three tunings are used to cover a contiguous coverage between 107~GHz, while for the 14~GHz, two tunings are sufficient to cover the observed wavelengths between 67 and 115~GHz \citep{Carpenter2023}. The Band~2 setup extends the accessible redshift space to lower redshifts compared to Band~3-only strategies, while simultaneously increasing the fraction of sources for which two or more lines are detected, directly resolving the redshift degeneracy without requiring follow-up observations. Band~2 enables a higher robust redshift fraction across both the HerBS and SPT redshift distributions.}
\label{fig:efficiencies}
\end{figure*}

\begin{table}
\caption{Assumed values of the ALMA receivers, current and in the WSU}
    \label{tab:WSUgoals}
    \centering
    \begin{tabular}{ccccccc}
    \hline
        Band &  RF Range  & IF Range & Type & IF Range & Type \\ \hline
             & & \multicolumn{2}{c}{\bf Current} & \multicolumn{2}{c}{\bf WSU} \\ \hline
        1 & 35-50    & 4-12   & SSB & 4-12 & SSB \\
        2 & 67-116   & 4-12      & 2SB & 2-18 & 2SB \\
        3 & 84-116   & 4-8    & 2SB & -    & -   \\
        4 & 125-163  & 4-8    & 2SB & 2-18 & 2SB \\
        5 & 163-211  & 4-8    & 2SB & 2-18 & 2SB \\
        6 & 209-281  & 4.5-10 & 2SB & 4-20 & 2SB \\
        7 & 275-373  & 4-8    & 2SB & 4-20 & 2SB \\
        8 & 385-500  & 4-8    & 2SB & 4-18 & 2SB \\
        9 & 602-720  & 4-12    & DSB & 4-20 & 2SB \\
        10 & 787-950 & 4-12    & DSB & 4-20 & 2SB \\\hline
    \end{tabular}
\end{table}

\section{Prospects of redshift surveys in the Band 2 and WSU era}
\label{sec:zsearches}
Identifying the cosmological redshifts of SMGs/DSFGs has long been challenging. With the completion of the full spectral coverage of ALMA, I will now use the public optimisation tool reported in \cite{Bakx2022}\footnote{The code is publicly available at \url{https://github.com/tjlcbakx/redshift-search-graphs}} to evaluate and optimize the ability of present-day and future ALMA at providing robust spectroscopic redshifts. 

The large beams of single-dish surveys hamper optical/near-infrared counterpart identification, while the heavy dust obscuration renders these galaxies faint at shorter wavelengths \citep[e.g.,][]{daCunha2015}. Radio-based astrometry offered only partial relief, given the large scatter in the far-infrared-to-radio luminosity ratio and the rapid decline in radio flux at $z \gtrsim 3.5$. 
These limitations drove the development of wide-bandwidth spectroscopic instruments in the submillimetre. Early instruments achieved relative bandwidths $\delta f/f < 1$\%, but successive generations, i.e., Z-Spec \citep{Naylor2003}, the Redshift Search Receiver \citep{Erickson2007}, Zpectrometer \citep{Harris2007}, and EMIR \citep{Carter2012}, pushed this above 10\%, enabling less biased searches for CO emission lines. The first blind CO-based redshifts followed, using EMIR \citep{Weiss2009}, Zpectrometer \citep{Frayer2010}, Z-Spec \citep{Lupu2012}, and the RSR \citep{Zavala2015}. At slightly shorter wavelengths, atomic and ionic lines, particularly [C\,{\sc ii}], which can be $\sim$4000 times more luminous than CO(1--0) \citep{Stacey2010}, proved highly efficient redshift indicators, but due to atmospheric windows remain limited to high-redshift targets.

ALMA transformed the field by confirming robust redshifts for 23 SPT-selected DSFGs through blind line detections \citep{Vieira2013}. The initial approach covered the Band~3 atmospheric window with five contiguous tunings spanning 86--116\,GHz. 
This tuning offers an 83\% probability of detecting at least one spectral line for an SPT-like redshift distribution. 
However, a single detected line is insufficient to resolve the redshift degeneracy without follow-up observations, unless the source lies at $z > 4$ where multiple lines fall within the band. The Band~3-only setup also suffers from an inherent efficiency loss, where the 7.5\,GHz Intermediate Frequency overlap means the central spectral region is covered twice, i.e., by the Lower Side Band of higher-frequency tunings and the Upper Side Band of lower-frequency tunings. It is worth noting that, beyond their primary redshift-search purpose, these deeper spectral scans have proven valuable for detecting fainter lines such as atomic carbon (\ci{}) and for constructing composite spectra.

Two strategies have since been adopted to improve efficiency. First, ALMA introduced the ``spectral scan'' observing mode in recent Cycles, consolidating multiple frequency tunings into a single calibration batch and substantially reducing slew times and associated overheads. Second, a Band~3$+$4 (3 and 2\,mm) strategy proposed in \cite{Bakx2022} and implemented in \cite{Urquhart2022} delivers a markedly higher redshift recovery rate: 65\% of targeted sources are expected to yield robust redshifts, with an observed fraction of 73\%, far exceeding the 12\% predicted from Band~3 alone. NOEMA's PolyFIX correlator, providing $\sim$16\,GHz of instantaneous bandwidth, has demonstrated comparable efficiency \citep{fudamoto2017, neri2020, Cox2023}. 

In the coming years, ALMA's capabilities will be significantly enhanced by the Wideband Sensitivity Upgrade \citep[WSU;][]{Carpenter2023}, comprising both correlator and receiver upgrades. The new correlator will reduce observation times by up to 40 per cent, while the receiver upgrade is expected to double or quadruple present-day bandwidths, approaching the capabilities of NOEMA (Table~\ref{tab:WSUgoals}; from \citealt{Carpenter2023}). Together, these improvements will strongly benefit redshift searches \citep{Carpenter2023} by providing novel frequency ranges and larger instantaneous bandwidths.

Figure~\ref{fig:efficiencies} shows the redshift identification probability as a function of redshift for three ALMA configurations: the filled Band~3 tuning \citep{Weiss2013} as a baseline, and compares them to two set-ups using the new Band~2 receiver in the WSU era. Evaluated against the HerBS \citep{bakx18,Bakx2020Erratum} and SPT \citep{reuter20} redshift distributions, the comparison demonstrates that incorporating Band~2 substantially increases the fraction of sources yielding two or more spectral lines, enabling unambiguous redshift identification without follow-up observations across a wider range of redshifts.
As the WSU remains in active development, several specifications are not yet finalised. Here I explore two candidate instantaneous sideband bandwidths, 8 and 14~GHz, corresponding to the baseline and an optimistic goals, respectively. {\color{referee} For the 8~GHz bandwidth scenario, three tunings are used to cover a contiguous coverage between 107~GHz with one overlapping spectral window between 83 and 91~GHz, while for the 14~GHz bandwidth scenario, two tunings are sufficient to cover the observed wavelengths between 67 and 115~GHz \citep{Carpenter2023}. For the present-day 3.875 and potential future 8~GHz bandwidths, this is further improved by the wider IF range for Band~2. This allows the spectral windows to be put side by side contiguously, facilitating more efficient redshift searches without the overlapping spectral windows in the current filled Band~3 observing set-up.
}

{\color{referee2}
The frequency coverage of Band 2 is not the only consideration in redshift searches. In Figure~\ref{fig:efficiency}, I compare the performance of Band~2 against that of Band 4 for a fiducial dusty star-forming galaxy (DSFG) with an infrared luminosity of $L_{\text{IR}} = 10^{12}~L_{\odot}$. I assume the line brightnesses following the CO spectral line energy distribution (SLED) scaling from \citet{Hagimoto2023} for transitions from CO(1--0) up to CO(7--6). This is compared against the expected observation times based on the analytical equation from the ALMA Technical Handbook\footnote{\url{https://almascience.eso.org/proposing/technical-handbook}}. 

The $J > 2$ transitions are generally detected faster, and thus favour Band 4 at redshifts below $z < 3$, as can be seen in the bottom panel that shows the ratio of on-source observing time between Band 2 and Band 4.
However, Band 2 covers a larger redshift per unit of frequency. As a comparison, the redshift range covered by Band~2 is compared to Band~4 with a solid black line, i.e., the d$z_{\rm Band 2}$/d$z_{\rm Band 4}$. While both receivers cover 8~GHz, Band~2 typically covers between 50 to 100~per cent larger redshift regions. An observing set-up ensuring a contiguous coverage of Band~4 furthermore overlaps two to three of the upper side-band of the lower frequency tunings and the lower side-band of upper frequency tunings, resulting in a further reduction of the observing efficiency of 20 to 25~per cent.
Regardless, this still implies that Band~4 is the better choice for most redshift searches in the $z < 3$ Universe, especially considering Band~4 probes a brighter region of the dust continuum, providing accurate astrometry to extract the line fluxes \citep[e.g.,][]{Bendo2023}. That said, for surveys Band~2 is set to become the premier choice due to the superior instantaneous bandwidth and the ability to contiguously stack spectral windows without the 4--8~GHz frequency gaps inherent in higher-frequency receivers. I find that Band 2 becomes significantly more efficient at $z > 3$, where the ability to cover a continuous frequency range without the ``ladder''-like tuning overlaps currently required in Bands 3 and 4 outweighs the lower line fluxes. For large-volume blind surveys, Band 2 thus provides a more robust and time-efficient solution for spectroscopic redshift identification.

\begin{figure}
    \centering
    \includegraphics[width=\columnwidth]{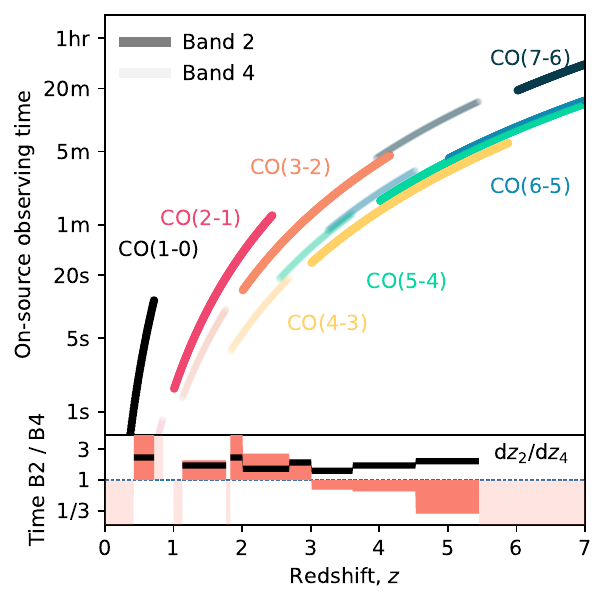}
    \caption{ \color{referee2} Observational efficiency comparison between ALMA Band 2 and Band 4 for a fiducial $L_{\text{IR}} = 10^{12}~L_{\odot}$ galaxy. The \textit{top panel} shows the required on-source observing time to achieve a $5\sigma$ detection for CO transitions across redshifts $z=0-7$, assuming line brightness conversions from \citet{Hagimoto2023}. The \textit{bottom panel} shows the ratio of on-source observing time ($t_{\text{B2}} / t_{\text{B4}}$). A ratio below 1 indicates that Band 2 is more efficient. The \textit{black solid line} quantifies the enhanced redshift coverage at lower frequencies, i.e., the d$z_{\rm Band 2}$/d$z_{\rm Band 4}$. The wide IF range of Band 2 allows for contiguous stacking, making it highly competitive for high-redshift line searches despite the required longer integrations compared to Band 4.}
    \label{fig:efficiency}
\end{figure}
}

\begin{figure}
    \centering
    \includegraphics[width=\linewidth]{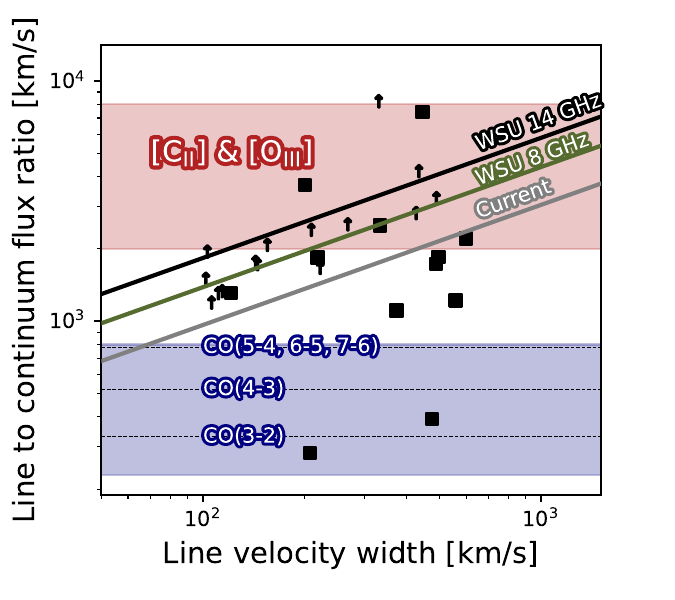}
    \caption{The equivalent width of an ALMA-detected spectral line can serve as a diagnostic, particularly in the WSU era. The velocity-integrated line flux compared to the adjacent dust continuum is much larger for \ciil{} and \oiiil{} emission \citep{Bakx2024GoldenRatio}, and thus blindly-found emission lines without associated dust continuum are likely from atomic fine-structure lines. Detections and upper-limits of emission line galaxies (\textit{black squares and arrows}) are from \citet{Venemans2020} and \citet{Fudamoto2021}, while the filled regions are from scaling relations by \citet{delooze14} and \citet{Hagimoto2023}. The diagnostic capabilities of ALMA towards finding such line emitters will improve as larger bandwidths become available in the WSU era. }
    \label{fig:dV_EW}
\end{figure}
Beyond DSFGs, ALMA's short-wavelength reach has enabled spectroscopic confirmations of UV-selected galaxies into the Epoch of Reionisation \citep{Inoue2016, Hashimoto2018, Tamura2019}. At these redshifts, Ly$\alpha$ is heavily attenuated due to the larger neutral fraction of the extragalactic medium and \textit{JWST} NIRSpec spectro-photometric redshifts become unreliable beyond $z \sim 10$, making submillimetre line searches the most robust path to redshift confirmation. 
Meanwhile, the decrease of noise temperatures towards longer wavelengths improves ALMA's sensitivity of lines at higher redshifts. With more galaxies being identified at higher redshift \citep[e.g.,][]{Carniani2024Natur}, a cosmologically-shifted \oiii{} emission line at $z = 20$ would only requires fifty per cent extra observing time compared to a similarly-bright line at $z = 14$ \citep{Carniani2025, Schouws2025}.
As mentioned briefly in \cite{Bakx2024GoldenRatio}, the larger bandwidths available in the WSU era will enable a larger parameter space to find line-of-sight galaxies with spectral line detections. Figure~\ref{fig:dV_EW} shows the increased capability of ALMA to distinguish true high-redshift galaxies from \cii{} and \oiii{}. Emission-line galaxies provide a search of massive galaxies without a prior UV selection, and are thus able to probe colder, dust-obscured galaxies \citep{Walter2018,Venemans2020,Fudamoto2021}. With detections of \oiiil{} now possible at $z > 12$ \citep[e.g.][]{Zavala2024,Carniani2025,Schouws2025}, these offer an opportunity to reveal nearby dust-obscured galaxies at a different stage in their evolution \citep{Ferrara2025} as an alternative to targeting dust-obscured galaxies with direct \textit{JWST} characterization \citep{Donnan2025,Mitsuhashi2025,Rodighiero2026}.

\section{Conclusions}
\label{sec:conclusions}

I have presented an analysis of ALMA Band~2 Science Verification observations of the strongly lensed DSFG SPT0027-50 at $z = \zspec$, exploiting for the first time ALMA's complete spectral coverage in the 3--4\,mm atmospheric window. Our main findings are as follows:

\begin{enumerate}

\item The Band~2 spectral scan detects eight emission lines, including CO(3--2), CO(4--3), \ci{} 609\,$\mu$m, and four dense-gas tracers (HCN(5--4), HNC(4--3), HNC(5--4), HCO$^{+}$(5--4)), providing a contiguous view of the molecular and dense gas inventory of a high-redshift DSFG. The CO(3--2)/CO(4--3) flux ratio of $1.7 \pm 0.3$ approaches the thermalization limit, indicating highly excited, dense gas typical of vigorously star-forming systems at Cosmic Noon.

\item The \ci{}/CO(4--3) line luminosity ratio of $0.30 \pm 0.06$ implies gas densities of $n_{\rm H} \sim 10^{4.5}$\,cm$^{-3}$ and a far-ultraviolet radiation intensity of $10^{3.8}$\,Habing, consistent with an intense starburst environment. The estimated dense gas fraction of $17 \pm 9$\,per\,cent is in agreement with values found for other DSFGs selected from large-area surveys.

\item High-resolution Band~7 continuum imaging reveals a clumpy lensed morphology in which star-forming clumps contribute 30--50\,per\,cent of the total emission, resolving structures on scales of $\sim$150\,pc. This places SPT0027-50 within the broader class of gas-rich, clumpy starburst systems at $z \sim 3$--$4$, and motivates future high-resolution molecular line follow-up to determine the dynamical origin of the clumps.

\item Looking ahead to the WSU era, Band~2 configurations with 8 and 14\,GHz instantaneous bandwidths substantially outperform the Band~3-only baseline for blind redshift searches, extending the accessible redshift space to lower redshifts and increasing the fraction of sources for which two or more spectral lines are simultaneously detected. This enables unambiguous redshift identification {\color{referee} for (sub)millimetre-selected sources} without the need for follow-up observations regardless of the redshift distribution of the sample.

\end{enumerate}

The observations reported here demonstrate that ALMA Band~2 is a powerful new window for ISM characterisation and redshift confirmation of high-redshift DSFGs. As integration times increase and the WSU upgrades are commissioned, Band~2 will become a cornerstone tool for spectroscopic surveys of the obscured Universe.

\section*{Acknowledgements}
{\color{referee}TB kindly thanks the anonymous referee and editors at MNRAS for their insightful comments and suggestions to improve this manuscript.}
TB acknowledges financial support from the Knut and Alice Wallenberg foundation through grant no. KAW 2020.0081.
The author kindly thanks fruitful discussions with Carlos De Breuck, Elvire De Beck, Kiana Kade, Patrick Kamieneski, Kirsten Knudsen, Louise Paquereau and Maria D\'iaz-Trigo. 
This paper makes use of the following ALMA data: ADS/JAO.ALMA\#2011.0.00025.SV, 
\#2015.1.00504.S,
\#2022.1.00172.S,
\#2022.1.00526.S, and
\#2023.1.01585.S. ALMA is a partnership of ESO (representing its member states), NSF (USA) and NINS (Japan), together with NRC (Canada), NSTC and ASIAA (Taiwan), and KASI (Republic of Korea), in cooperation with the Republic of Chile. The Joint ALMA Observatory is operated by ESO, AUI/NRAO and NAOJ.

\section*{Data Availability}
This project makes use of the tool described in \cite{Bakx2022} that is also available publically at \url{github.com/tjlcbakx/redshift-search-graphs}.
All observational data is available on the ALMA Science Archive, and is reduced using the scripts provided.


\bibliographystyle{mnras}
\bibliography{example}








\bsp 
\label{lastpage}
\end{document}